• Article •

# The first result on $^{76}$Ge neutrinoless double beta decay from CDEX-1 experiment

Wang Li[1], Yue Qian[1*], Kang KeJun[1], Cheng JianPing[1], Li YuanJing[1], Wong TszKing Henry[7†], Lin ShinTed[2,7], Chang JianPing[4], Chen JingHan[7†]  Chen QingHao[1], Chen YunHua[6], Deng Zhi[1], Du Qiang[4], Gong Hui[1], He Li[5], He QingJu[1], Hu JinWei[1], Huang HanXiong[2], Huang TengRui[7†], Jia LiPing[1], Jiang Hao[1], Li HauBin[7†], Li Hong[5], Li JianMin[1], Li Jin[1], Li Jun[5], Li Xia[3], Li XinYing[4], Li XueQian[4], Li YuLan[1], Lin FongKay[7†], Liu ShuKui[1], Ma Hao[1], Ma JingLu[1], Pan XingYu[1], Ren Jie[3], Ruan XiChao[3], Shen ManBin[6], Sharma Vivek[7,8†], Singh Lakhwinder [7,8†], Singh Manoj Kumar [7,8†], Singh Manoj Kumar [7,8†], Soma Arun Kumar [7,8†], Su Jian[1], Tang ChangJian[4], Tang WeiYou[1], Tseng ChaoHsiung[7†], Wang JiMin[6], Wang Qing[1], Wu ShiYong[6], Wu Wei[1], Wu YuCheng[1], Xing HaoYang[2], Xu Yin[4], Xue Tao[1], Yang LiTao[1], Yang SongWei[7†], Yi Nan[1], Yu ChunXu[4], Yu HaiJun[5], Zeng WeiHe[1], Zeng XiongHui[6], Zeng Zhi[1], Zhang Lan[5], Zhang YunHua[6], Zhao MingGang[4], Zhao Wei[1], Zhong SuNing[4], Zhou JiFang[6], Zhou ZuYing[3], Zhu JingJun[2], Zhu WeiBin[5], and Zhu ZhongHua[6]

(CDEX Collaboration)

$^{1}$Key Laboratory of Particle and Radiation Imaging (Ministry of Education)
and Department of Engineering Physics, Tsinghua University, Beijing 100084
$^{2}$College of Physical Science and Technology, Sichuan University, Chengdu 610064
$^{3}$Department of Nuclear Physics, China Institute of Atomic Energy, Beijing 102413
$^{4}$School of Physics, Nankai University, Tianjin 300071
$^{5}$NUCTECH Company, Beijing 10084
$^{6}$YaLong River Hydropower Development Company, Chengdu 610051
$^{7}$Institute of Physics, Academia Sinica, Taipei 11529
$^{8}$Department of Physics, Banaras Hindu University, Varanasi 221005

We report the first result on $^{76}$Ge neutrinoless double beta decay from CDEX-1 experiment at China Jinping Underground Laboratory. A mass of 994 g p-type point-contact high purity germanium detector has been installed to search the neutrinoless double beta decay events, as well as to directly detect dark matter particles. An exposure of 304 kg·day has been analyzed. The wideband spectrum from 500 keV to 3 MeV was obtained and the average event rate at the 2.039 MeV energy range is about 0.012 count per keV per kg per day. The half-life of $^{76}$Ge neutrinoless double beta decay has been derived based on this result as: $T^{0\nu}_{1/2} > 6.4 \times 10^{22}$ yr (90% C.L.). An upper limit on the effective Majorana-neutrino mass of 5.0 eV has been achieved. The possible methods to further decrease the background level have been discussed and will be pursued in the next stage of CDEX experiment.

neutrinoless double beta decay, CDEX, point-contact germanium detector, $^{76}$Ge isotope

* Corresponding author (email: yueq@mail.tsinghua.edu.cn)
† Participating as a member of TEXONO Collaboration.

## 1 Introduction

The atmospheric and solar neutrino oscillation experiments such as Super-Kamiokande and SNO have confirmed neutrino oscillations each other and thus at least two of the three neutrinos have non-zero masses, if not all three [1, 2, 3]. The differences of the squared mass eigenstates of three neutrinos has been measured though the absolute mass scale is unknown until now. So the neutrino mass hierarchy problem has been a hot topic in the neutrino domain, together with other important problems [4, 5, 6]. Neutrinoless double beta decay ($0\nu\beta\beta$) experiment has been one of the important experiments due to its strong relation with several important topics including the confirmation of the Dirac or Majorana character of neutrino, the lepton number conservation problem, and so on [7].

Neutrinoless double beta decay can occur as following mode if the neutrino is its own anti-particle, i.e the neutrino is a Majorana particle: $(A, Z) \rightarrow (A, Z+2) + 2e^-$. It is very critical because this process has violated the lepton number conservation. At the same time, many new physics beyond the standard model could be studied based on the mass of the Majorana-neutrino.

The two electrons take all of the energy of the neutrinoless double beta decay process and will contribute a peak at the Q-value ($Q_{\beta\beta}$) position of the $0\nu\beta\beta$ decay spectrum.

In principle, the $0\nu\beta\beta$ decay could only occur for the even-even nuclei due to the transition forbidden. After long time research, scientists have picked up several nuclei which are quite suitable to be the target nuclei for $0\nu\beta\beta$ and listed as follow: $^{48}$Ca, $^{76}$Ge, $^{82}$Se, $^{100}$Mo, $^{130}$Te, $^{136}$Xe, $^{150}$Nd, and so on.

There are many experiments to search the neutrinoless double decay events from different nuclei based on different detector technologies such as GERDA[8], MAJORANA[9], EXO [10], CUORE [11], NEMO [12], SNO+ [13], EXO [14].

Due to its very good energy resolution, very low internal background, and easy scale-up, high purity germanium (HPGe) detector has been one of the most competitive technologies for $0\nu\beta\beta$ experiment served as target nuclei and detector at the same time. The Heidelberg-Moscow experiment had run five 86% enriched $^{76}$Ge detectors with totally active mass of 10.96 kg to achieve a $^{76}$Ge $0\nu\beta\beta$ decay half-life of $T^{0\nu}_{1/2} > 7.4 \times 10^{24}$ year (90% C.L.) [15]. The first result from IGEX has been published in 1999 and the lower bound is $T^{0\nu}_{1/2} > 0.8 \times 10^{25}$ year (90% C.L.) [16]. A new collaboration, GERDA, has been setup in 2004 partially based on the enriched $^{76}$Ge detectors developed by Heidelberg-Moscow and IGEX groups. GERDA collaboration has now setup an array of HPGe detectors with total mass of less than 40 kg which have been run in Gran Sasso underground laboratory, Italy. The HPGe detector array is cooled and active shielded with liquid argon to further lower its background level. The best result for the half-life of $^{76}$Ge isotope has been published by GERDA collaboration in 2013 as follow: $T^{0\nu}_{1/2} > 2.1 \times 10^{25}$ yr (90% C.L.) [8].

Majorana Demonstrator, which is the first stage of the Majorana collaboration, has also setup an array HPGe detector system with less than 40 kg total mass for double beta decay search in Sanford underground laboratory at Homestake, USA. The results are under preparation for the half-life measurement of $^{76}$Ge $0\nu\beta\beta$ decay [9].

CDEX collaboration is pursuing the direct detection of light dark matter with mass of less than 10 GeV and the studies of double beta decay in $^{76}$Ge with ton-scale point-contact germanium detector at the China Jinping Underground Laboraotry (CJPL) [17, 18, 19]. The CDEX-1 experiment, the first stage of CDEX collaboration in order to study the performances of the CDEX prototype detectors, has developed a 1kg-scale p-type point-contact germanium detector with several output channels to cover both low energy region (down to 500 eV) for dark matter detection and high energy region (up to 3 MeV) for double beta decay experiment [20]. The physical results for both Weakly Interactive Massive Particles (WIMPs, donated as $\chi$) and Axion dark matter detection have been published based on different dataset from 2013 on [21, 22, 23, 24]. The most stringent sensitivity for WIMP-nucleus spin-independent interaction based on a point-contact germanium detector has been achieved by CDEX-1 experiment and the favored region by CoGeNT experiment has been excluded with an identical detector technique [23]. The measured cross sections of spin-dependent $\chi$-neutron were also published and it is the best sensitivities below 6 GeV/$c^2$ [24]. The CDEX-1 limits on galactic dark matter axion has been the most stringent at axion mass of less than 1 keV [25].

We report in this paper the half-life results for $^{76}$Ge neutrinoless double beta decay and the consequent upper limit on the effective Majorana-neutrino mass based on the dataset of CDEX-1 experiment.

## 2 Experimental setup

The CDEX-1 detector system was installed in China Jinping Underground Laboratory which locates at Jinping traffic tunnel, Sichuan province, China and has a rock overburden of 2400 m. The muon flux in CJPL is about 61.7 yr$^{-1}$ m$^{-2}$ [26] and the ambient gamma radioactivity is measured to be about $3.94 \times 10^3$ kg$^{-1}$ keV$^{-1}$ day$^{-1}$ based on an *in situ* HPGe detector [27]. The fast neutron and thermal neutron fluxes are $4.00 \times 10^{-6}$ cm$^2$ s$^{-1}$ [28] and $1.5 \times 10^{-7}$ cm$^2$ s$^{-1}$ [29] measured with Gd-doped liquid scintillator and $^3$He proportional chamber, respectively. Due to its deepest rock overburden, quite pure ambient rock, and direct access by car, CJPL has been a perfect underground laboratory suitable for the rare event experiments such as dark matter direct detection, neutrinoless double beta decay and solar neutrino experiment, and so on. The dimension of the main hall of CJPL phase-I is about 6.5 m (Height) × 6.5 m (Width) × 40 m (Length) which CDEX-1 detector system is located in.

A polyethylene room with wall of 1m-depth has been constructed in CJPL main hall to provide a space with dimension of 4 m (Height) × 4 m (Width) × 8 m (Length) inside to house the CDEX-1 detector system. The CDEX-1 point-contact HPGe detector was shielded passively from outside to inside in the polyethylene room including 20 cm lead, 20 cm Boron-doped polyethylene, 20 cm oxygen free high conductivity (OFHC) copper. Inside the OFHC copper shielding, there is a low background NaI(Tl) anti-Compton active shielding detector to enclose the cryostat of CDEX-1 HPGe detector. The detailed shielding structure could be referenced in [24].

The CDEX-1 detector is a prototype single-module HPGe detector fabricated with a p-type germanium crystal of 994 g of mass which dimension is about diameter of 62 mm and height of 62 mm. The CDEX-1 HPGe detector worked at +3500 V provided by a high voltage module (CANBERRA 3106D). One output from the tiny P+ electrode was fed into a Canberra 2111 fast amplifier in order to cover a wide dynamic energy range of up to 3 MeV for $0\nu\beta\beta$ decay experiment, together with several outputs to be specially tuned to favor the low energy region below 1 keV for dark matter detection. The output from NaI(Tl) anti-Compton detector, and the inhibit signal from the CDEX-1 point-contact HPGe detector were recorded at the same time for offline data analysis. These outputs were digitized and recorded by a flash analog-to-digital convertor (FADC, CAEN V1724) at a 100 MHz sampling rate with a resolution of 14 bits. The data acquisition software is based on the LabVIEW program. Random trigger signals (RT) at 0.05 Hz generated by a precision pulser were injected into the system, in order to monitor the noise level and dead time of the data acquisition system and provide the system trigger. The total DAQ trigger rate is about 3~5 Hz in order to decrease the penalty of dead time. The detailed structure of the CDEX-1 HPGe detector and the schematic diagram of the data acquisition system could be found in reference [20] in details.

The high purity germanium crystal of CDEX-1 detector is natural germanium, of which the faction of $^{76}$Ge atoms is 7.83%. The detector has an approximately 1.02±0.14 mm dead layer in the top and lateral side of the crystal, which give rise to a fiducial mass of 915g with an uncertainty of 1.0% [30]. Datasets from Nov, 2013 to Dec, 2014 have been used for the $0\nu\beta\beta$ decay half-life evaluation, corresponding to a total exposure of 304 kg·day live data or 0.898 mol·yr of $^{76}$Ge.

## 3   Data analysis

The digitized charge pulse from the high energy output channel was parameterized based on the data analysis methods and procedures established in previous study of CDEX-1 experiment [21]. Time information and pulse shape characteristics of each output signal were extracted from the original charge pulse. The integrated pulse area of one original signal was used to denote the energy of this event. To get the best energy resolution, the integrating range is optimized according to the Full Width Half Maximum (FWHM) of the characteristic gamma peaks in the high energy range, such as 1.46 MeV peak from $^{40}$K, 1.76 MeV peak from $^{214}$Bi, and 2.61 MeV from $^{208}$Tl.

The total data was divided into 35 datasets in terms of the nitrogen filling period. It has been found that there is a slightly gain fluctuation between different datasets. Two datasets were abandoned due to their relatively bad energy resolution (FWHM), caused by abnormal gain fluctuations during the dataset acquisition.

Anti-coincidence (AC) cut was applied in the data selection procedure. According to the trigger time difference of the anti-Compton detector and the HPGe detector, anti-coincidence events were selected and excluded. As independent with energy, the efficiency of the AC cut was measured by the random trigger events as is approximately 100%.

Using the characteristic gamma peaks from natural radioactivities of the detector itself, independent calibrations were performed for each dataset. Figure 1 shows the calibration curve of one dataset. Five gamma peaks from $^{214}$Bi (609.3 keV, 1.764 MeV), $^{228}$Ac (911.2 keV), $^{40}$K (1.461 MeV) and $^{208}$Tl (2.615 MeV) has been used, and the calibration curve was obtained by fitting these points with a second-order polynomial. The calibration results show that the deviation of reconstructed peak positions of the gamma lines was smaller than 0.3 keV to their nominal values.

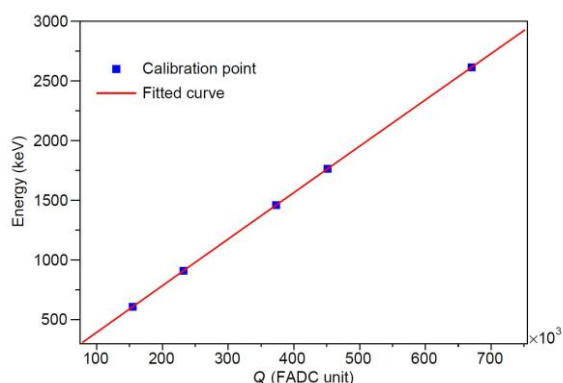

**Figure 1**   Energy calibration of one of the 35 datasets.

A typical pulse from the P+ electrode channel of the CDEX-1 PCGe detector was parameterized with several numbers to describe its characteristics as displayed in Figure 4 of reference [24]. Two energy-related parameters are defined: (i) the maximal amplitude of one pulse ($A_{max}$) and (ii) the integration of one pulse which was chosen to define as energy (E) for its excellent energy resolution and linearity at the high-energy range. The germanium detector has ability of single-site event (SSE) and multi-site event (MSE) discrimination, based on the pulse shapes of the charge signals. A parameter of $A_{max}/E$ was used to discriminate the events of SSE and MSE as shown in Figure 2, SSE has a relatively large pulse height than MSE in the case of same energy events. $0\nu\beta\beta$ decay event is mainly SSE in bulk volume of the crystal, resulting in a nearly Gaussian distribution of the $A_{max}/E$. The energy resolution of the detector for SSEs can be extracted from SSE cut. The cut lines are determined by fitting the $A_{max}/E$ distribution boundaries of SSE of the Compton continuum parts in the energy region of larger than 1MeV in Figure 2. The whole datasets have been divided into 4 groups in chronological order and an independent analysis has been performed for each group. Figure 3 shows the fitting line of the energy resolution of the SSEs for one of them. The FWHM of gamma peaks from $^{214}$Bi (609.3 keV, 1.764 MeV), $^{228}$Ac (911.2 keV) and $^{40}$K (1.461 MeV) has been fitted with a function $FWHM = a + b \cdot \sqrt{E}$ and the average extrapolated FWHM at $Q_{\beta\beta}$ is 4.3±0.2 keV. In fact we just use the SSE cut to calculate the energy resolution for SSE events at $Q_{\beta\beta}$ in this paper and do not apply this SSE cut to further decrease the background level of the spectrum due to the quite complicated efficiency correction. This will be tried in the next stage of CDEX double beta decay experiment.

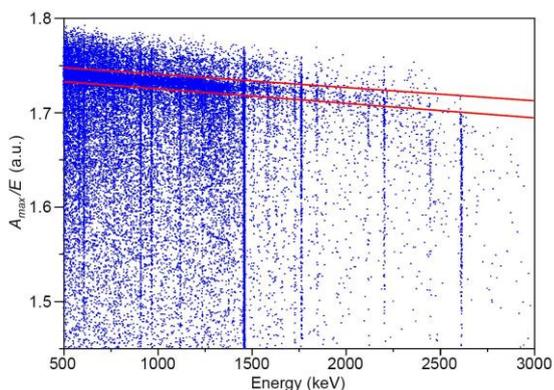

**Figure 2** (Color online) SSE and MSE discrimination. The red curves are the upper and lower cut lines for SSEs.

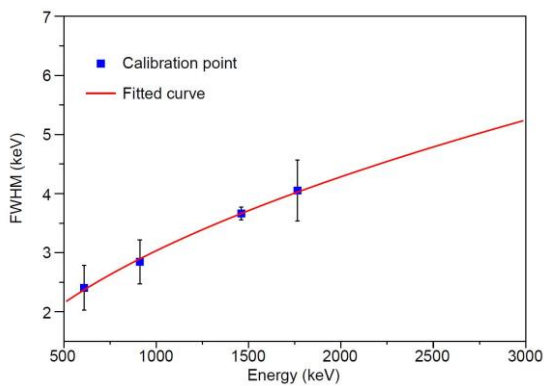

**Figure 3** Energy resolution of the single-site events (SSEs).

The preamplifier used in CDEX-1 is a pulsed reset preamplifier which works in a pulsed mode and it was reset to its initial condition by discharging the FET quickly marked by one Reset Inhibit Signal [24]. As shown in Figure 4(a), the charge of one event which has a high energy and locates close to the end of the reset period may not be completely collected. This kind of events can be excluded by the reset inhibit signal, however, it will induce an energy dependent efficiency. If there is only one physical event in a reset period, the efficiency can be calculated as:

$$\eta(E) = \frac{E_s - E_s \dfrac{20ms}{T_0} - E}{E_s - E_s \dfrac{20ms}{T_0}} \qquad (1a)$$

$$\frac{1}{T_0} = \frac{1}{n} \sum\nolimits_{i=1}^{n} \frac{1}{T_{0i}} \qquad (1b)$$

$E_s$ is the saturation energy of the preamplifier, E is the energy of this event. $T_{0i}$ is the reset period if there is no signal in this period. The reset period changes between 600 ms and 1200 ms during the whole data acquisition time. $1/T_0$ is the weighted average of $1/T_{0i}$. 20ms means a time period after the reset signal, within which the FADC will be vetoed to avoid the noises induced by the Reset Inhibit Signal. Figure 4(b) shows the calculated efficiencies with the energies in the interested range and the efficiency at $Q_{\beta\beta}$ is 78.7%. Due to the quite low background rate, the multi-event situation in one reset period is negligible. The artificial dead time due to the 20ms after the reset signal accounts for 2.1% of the total runtime. This caused an additional efficiency of 97.9%.

The energy spectrum obtained according to the AC cut and the efficiency corrections is shown in Figure 5 and the main gamma peaks were identified and labeled in the same plot. The background level in $Q_{\beta\beta}$ energy range is about 0.012 counts/(kg keV day). Note that the SSE cut has not been used for background suppression due to its unknown discrimination efficiency.

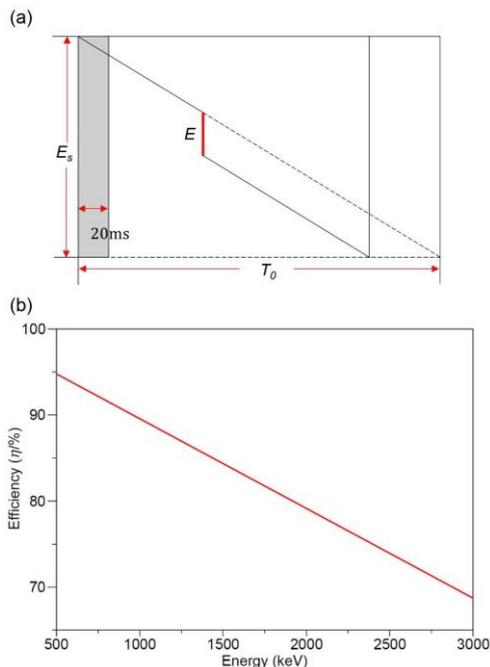

**Figure 4** (a) The principle of the efficiency change for the different events with different energies due to the Reset Inhibit Signal produced by the pulsed reset preamplifier of the CDEX-1 detector and (b) the efficiencies with the related energies in our interested energy range from 500 keV to 3 MeV.

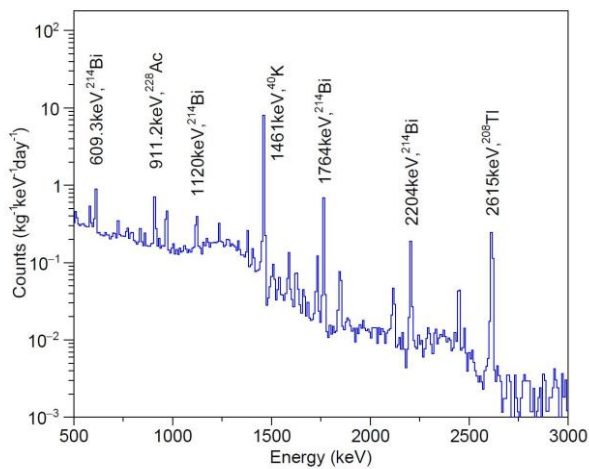

**Figure 5** Spectrum of 304 kg·day data from CDEX-1. The main gamma lines were labeled.

## 4 Background model

In order to understand the background contribution in $Q_{\beta\beta}$ energy range of the CDEX-1 detector, simulations spectra of different radioactive impurities in the different locations of the detector setup were performed based on Geant 4.10.2. A dead layer of 1.02±0.14 mm in the top and lateral side of the germanium crystal of the CDEX-1 detector is set in the simulation. Table 1 lists all the simulated background components and their locations. In the simulation, $^{238}$U decay chain and $^{232}$Th decay chain were treated as secular equilibrium and uniformly distribution. However, considering the radon gas contamination from outside space though pure nitrogen gas are flushing during the data taking, the $^{238}$U chain in copper shield is treated as two parts, one is the isotopes before $^{222}$Rn, and the other is the isotopes which begin from $^{222}$Rn in the decay chain. The simulated locations of the radioactive impurities are mainly in five parts of the experiment setup: the copper shield, NaI(Tl) crystal, the copper shell of the germanium detector, the electronics of the germanium detector and the germanium crystal itself. The simulated radioactive impurities include natural decay chains of $^{238}$U and $^{232}$Th, $^{40}$K, and these cosmogenic isotopes in germanium crystal and the detector copper shell. In addition, $^{40}$K decays in the PMT window of the NaI(Tl) Anti-Compton detector were considered in the simulation. Due to the extremely low muon flux in CJPL [26], muons and their secondary particles have not be considered. Neutrons are also negligible due to their quite low flux in the 1m-depth polyethylene room in CJPL main hall. Simulation shows that gamma induced energy spectra from sources in the electronics and the detector copper shell have similar shapes in high energy range. Their contributions can be represented by one of them in this analysis. And the background contribution of NaI(Tl) crystal is negligible because the vast majority events from it are excluded by the AC cut.

**Table 1** Simulated background components for CDEX-1 experiment

| Source | Location |
| --- | --- |
| cosmogenic isotopes | Ge Crystal |
| | Detector Copper Shell |
| $^{238}$U chain | Copper Shield |
| | NaI |
| | Detector Copper Shell |
| | Electronics |
| $^{232}$Th chain | Copper Shield |
| | NaI |
| | Detector Copper Shell |
| | Electronics |
| $^{40}$K | Copper Shield |
| | PMT Window |
| | NaI |
| | Detector Copper Shell |
| | Electronics |

Figure 6 shows the background model and its decomposition. The background model was obtained by fitting the experimental spectrum with the simulated spectra of all the components in the energy range of 500-3000 keV, using a least-squares method. The contribution of 2νββ was calculated on the basis of the half-life measured in reference [31]. Contributions of the cosmogenic isotopes (except $^{60}$Co) in the germanium crystal were determined by the corresponding Kx characteristic peaks in the low energy range of its experimental spectrum (0-12keV) [24]. Contributions of other radioactive impurities were determined by the best fit of the background model.

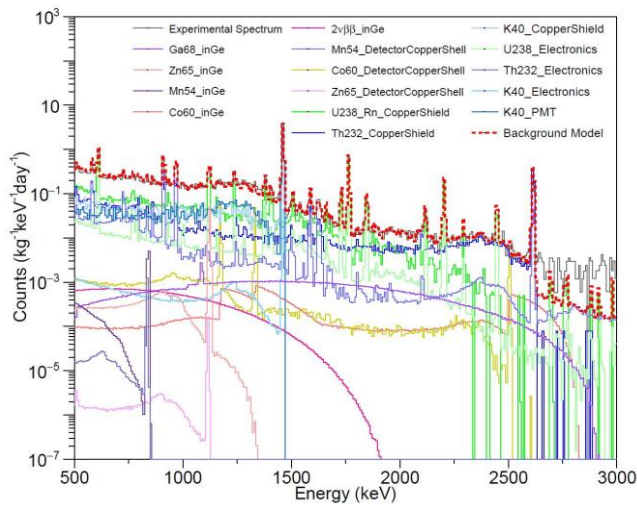

**Figure 6** (Color online) The decomposition of the background spectrum of CDEX-1 detector in the high energy of 500 keV to 3 MeV.

The background model shows that the main background contributions are from $^{232}$Th and $^{238}$U decay chain in the copper shield and the preamplifier electronics of the germanium detector. $^{68}$Ga and $^{60}$Co in the germanium crystal and $^{60}$Co in the copper shell also have a relatively high background contributions in the $Q_{\beta\beta}$ energy range. The background model predicts a flat background in the $Q_{\beta\beta}$ energy range which is used as a background model for the $0\nu\beta\beta$ event number estimation in the interested energy region.

## 5 Physics results

The total 304 kg·day data was used for the evaluation of $0\nu\beta\beta$ decay half-life of $^{76}$Ge. The backgrounds for different radioactivity sources have not been subtracted from the total spectrum and SSE/MSE discrimination has not been used to get rid of the MSE events due to its unknown discrimination efficiency in order to achieve a relatively conservative result.

The half-life of $0\nu\beta\beta$ decay of $^{76}$Ge isotope can be calculated as follows:

$$T_{1/2}^{0\nu} = \frac{\ln 2T \cdot M \cdot f_{76} \cdot N_A \cdot \varepsilon_1 \varepsilon_2 \cdot \varepsilon_3}{N^{0\nu} \cdot m} \quad (2)$$

where T is the measure time of the total data, M is mass of the detector, m is the molar mass of the natural germanium, $N_A$ is Avogadro's constant, and $f_{76}$ is the fraction of $^{76}$Ge atoms in natural germanium. $N^{0\nu}$ is the observed events of $0\nu\beta\beta$ decay. $\varepsilon_1$ is the detection efficiency for $0\nu\beta\beta$ events due to the reset period of the preamplifier. $\varepsilon_2$ is the efficiency due to the artificial dead time. The factor $\varepsilon_3$ represents the probability that the two electrons from one $0\nu\beta\beta$ decay event deposit all their energy in the active volume of the germanium detector, resulting in a full energy peak at 2039 keV, and its value is obtained by simulation. In the simulation, the energies of the two electron and the angle between their emitted directions are sampled on the basis of the distribution described by the theoretical formulas in reference [32]. The dead layer in the top and lateral side of the detector has also been considered in the simulation. The simulation result gives that the efficiency of the full energy deposit is (88.6±0.6) %.

A profile likelihood method has been used to evaluate the signal strength of $0\nu\beta\beta$ decay in the energy range 2020 keV - 2060 keV [33]. The fitted function consists of two parts: a constant term for background and a gauss function for $0\nu\beta\beta$ decay signals. The mean and the standard deviation of the gauss function is fixed according to the Q value of the $0\nu\beta\beta$ decay and the expected SSE energy resolution. The systematic uncertainties on the peak position (±0.3keV), the resolution and efficiencies have been considered by a Monte Carlo method. Figure 7 shows the energy spectrum without efficiency correction in this energy range. According to the fitting result, no peak was found at $Q_{\beta\beta}$ energy. The 90% C.L. limit of the half-life of $0\nu\beta\beta$ decay is

$$T_{1/2}^{0\nu} \geq 6.4 \times 10^{22} \ yr, \ 90\% \ C.L.$$

and the corresponding 90% C.L. upper limit of $0\nu\beta\beta$ signal strength is 4.0 events.

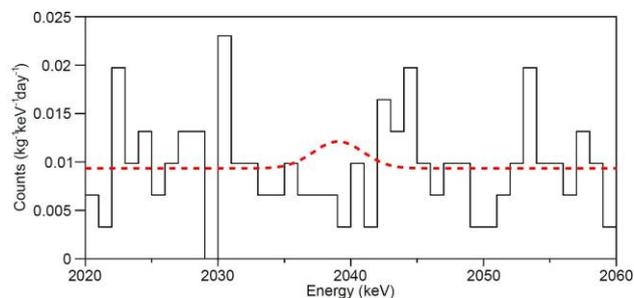

**Figure 7** Spectrum without efficiency correction in the region of interest for the $0\nu\beta\beta$ decay, the red dash line corresponds to 90% C.L. upper limit derived in this work superimposed the fitted background.

The measured $0\nu\beta\beta$ decay half-life limit of $^{76}$Ge gives an upper limit on the effective Majorana-neutrino mass of 5.0 eV. This result is obtained by using the nuclear matrix element (NME) from [34]. Figure 8 shows the comparison of this result with the limits from Gerda Collaboration phase I using the same NME [8]. The green and magenta areas in Figure 8 are the allowed values for the effective Majorana mass as a function of the smallest neutrino mass for both the inverted and normal mass hierarchies, according to the $1\sigma$ range values of neutrino oscillation parameters mentioned in reference [35].

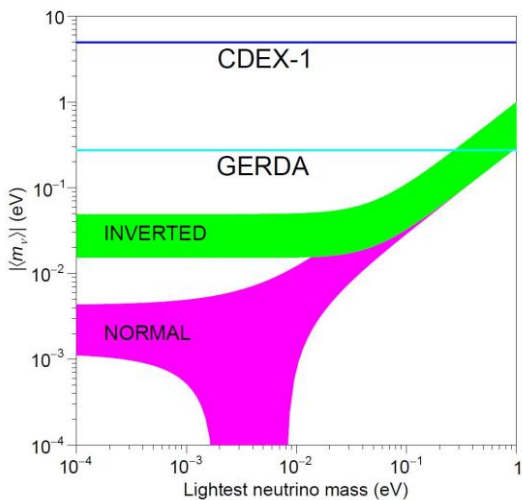

**Figure 8** Upper limits on the effective Majorana-neutrino mass from this work and Gerda Collaboration

# 6 Discussion and prospects

The first result about the half-life of $^{76}$Ge 0νββ decay has been measured with a 1kg-mass p-type point-contact natural germanium detector by CDEX Collaboration. The upper limit on the effective Majorana-neutrino mass of 5.0 eV from this work has been achieved. Due to its small $^{76}$Ge isotope mass, high background contributions from the preamplifier electronics and ambient materials in the vicinity of the detector, the background level in the $^{76}$Ge 0νββ decay energy range is about 400 times higher than that of GERDA-I experiment [8] which induces a relatively shorter half-life result for $^{76}$Ge 0νββ procedure.

The CDEX-1 PCGe detector has been designed by CDEX collaboration and fabricated by Canberra Company with up-to-date commercial low background structure materials and preamplifier parts. The higher background level shows that a better radioactive purification for the preamplifier, ambient structure and shielding materials near the detector should be prepared and selected by very sensitive technologies.

Several key technologies should be tackled in the next step to lower the background level of CDEX germanium detector: (1) use $^{76}$Ge-enriched crystal to increase the total number of $^{76}$Ge nucleus and to decrease further the cosmogenic background contributions from the long-life radioactive isotopes in the germanium crystal, such as $^{68}$Ge, $^{60}$Co and so on; (2) select ultra-pure materials to prepare the substrate of the preamplifier, the cables near the germanium crystal, the structure materials and the insulated materials in the vicinity of the germanium crystal, at the same time select the pure components such as capacitors and resistances for the preamplifier and use the materials as less as possible; (3) design more efficient passive and/or active shielding system with purer materials in order to reduce the background contributions from the outside circumstances. CDEX collaboration will focus on the development of the related technologies in the future to push the $^{76}$Ge 0νββ decay experiment forward.

*This work was supported by the National Natural Science Foundation of China (Grant Nos. 11275107, 11475117 and 11475099) and the National Basic Research Program of China (973 Program) (Grant No. 2010CB833006).*